\documentclass[aip,amsmath,amssymb,jap,reprint]{revtex4-1}
\usepackage{graphicx}
\usepackage{dcolumn}
\usepackage{bm}

\begin{document}

\title{Electronic structure of a graphene superlattice with massive Dirac fermions}

\author{Jonas R. F. Lima}

\address{Instituto de Ciencia de Materiales de Madrid (CSIC) - Cantoblanco, Madrid 28049, Spain}
\email{jonas.iasd@gmail.com}
\date{\today}

\begin{abstract}
We study the electronic and transport properties of a graphene-based superlattice theoretically by using an effective Dirac equation. The superlattice consists of a periodic potential applied on a single-layer graphene deposited on a substrate that opens an energy gap of $2\Delta$ in its electronic structure. We find that extra Dirac points appear in the electronic band structure under certain conditions, so it is possible to close the gap between the conduction and valence minibands. We show that the energy gap $E_g$ can be tuned in the range $0\leq E_g \leq 2\Delta$ by changing the periodic potential. We analyze the low energy electronic structure around the contact points and find that the effective Fermi velocity in very anisotropic and depends on the energy gap. We show that the extra Dirac points obtained here behave differently compared to previously studied systems.
\end{abstract}

\maketitle

\section{Introduction}

Graphene has attracted a great deal of attention since its first successful experimental fabrication  \cite{Novoselov} in 2004 due to its intriguing physics and application potential \cite{RevModPhys.81.109,RevModPhys.82.2673,RevModPhys.83.407}. Graphene is a one-atom thick layer of carbon atoms arranged in a hexagonal structure and its low-energy electronic structure can be described by using a Dirac-type Hamiltonian. The neutral, clean system has no gap and it is described by a massless Dirac equation. Due to the Klein tunneling \cite{Katsnelson,PhysRevB.73.241403}, charge carriers can not be confined by electrostatic potentials, what limits the uses of graphene in electronic devices. Opening a gap in the spectrum can help to confine the charges. 

An energy gap can be induced in graphene, for instance, by doping with boron \cite{bdg,bdg2} or nitrogen \cite{ndg} atoms. Another way to open an energy gap in the electronic structure of graphene is using an appropriate substrate. It was verified that a hexagonal boron nitride (\textit{h}-BN) substrate induces an energy gap of $53$ meV in graphene \cite{BN}, which can be tuned by transverse electric field \cite{Ilyasov}. Epitaxial graphene grown on SiC substrate has a gap of $\approx 0.26$ eV \cite{SiC}. The other electronic property of graphene that depends on substrate is the Fermi velocity \cite{Hwang}. The Klein tunneling can be suppressed also by electromagnetic fields \cite{Franco,Martino,Roy,Maksym} and by a spatially modulated gap \cite{peres,Giavaras,Lima201582,Lima2014}, leading to confined states.

In the last years, the possibility of engineering the electronic band structure of graphene by applying a periodic potential, i.e., a superlattice, has attracted considerable research interest to this subject. There are different methods to generate the periodic potential structure in graphene, such as electrostatic potentials \cite{Bai,Barbier2,Park,Peeters,Tiwari,Wang2,Barbier,Wang,PhysRevB.86.205422} and magnetic barriers \cite{Ramezani,Sankalpa,Vasilopoulos,Luca}. The combined effects of electrostatic and magnetic barriers have been studied as well \cite{Zhai,Moldovan}. Despite the difficulty of fabricating graphene under nanoscale periodic potentials, it was already realized experimentally \cite{Marchini,Calleja,Sutter,Martoccia,Rusponi,Yan}. It was found that the periodic potential leads to the appearance of extra Dirac points in the electronic structure of graphene \cite{Brey,Steven,Park,Wang,Wang2,Barbier,PhysRevB.86.205422} and affect the transport properties, inducing an anisotropy in the carriers group velocity \cite{Barbier,PhysRevB.86.205422}, leading to the collimation of electrons beams \cite{Marvin,PhysRevB.79.075123,Barbier}. The electronic structure of a bilayer and trilayer graphene superlattice were also analyzed \cite{Uddin}. Periodic potential can not open an energy gap in graphene.

In this paper, we investigate the electronic and transport properties of a graphene sheet deposited on a substrate that opens an energy gap in its electronic structure. On top of it we apply an external periodic potential. Our work is centered into analyzing the electronic structure in the vicinity of the new Dirac  points that arise by the interplay of the gap induced by the substrate and the driven periodic potential. We show that the gap can be tuned by the external periodic potential and that the new Dirac points show characteristic differences with respect to those found previously in similar systems. We analyze the electronic and transport properties in the vicinity of the contact points be obtaining the dispersion relation and the effective Fermi velocity, which turns out to be very anisotropic around the contact points and is sensitive to the energy gap.

The paper is organized as follows: In Sec. II we obtain the dispersion relation for the gapped graphene with a piecewise constant periodic potential. In Sec. III we investigate the electronic and transport properties of the system. We analyze the electronic band structure for equal and unequal well and barrier widths and investigate the emergency of extra Dirac points. We also find the dispersion relation and the group velocity around the contact points. The paper is summarized and concluded in Sec. IV.

\section{The Dispersion Relation}

\begin{figure}[hpt]
\centering
\includegraphics[width=8cm,height=5cm]{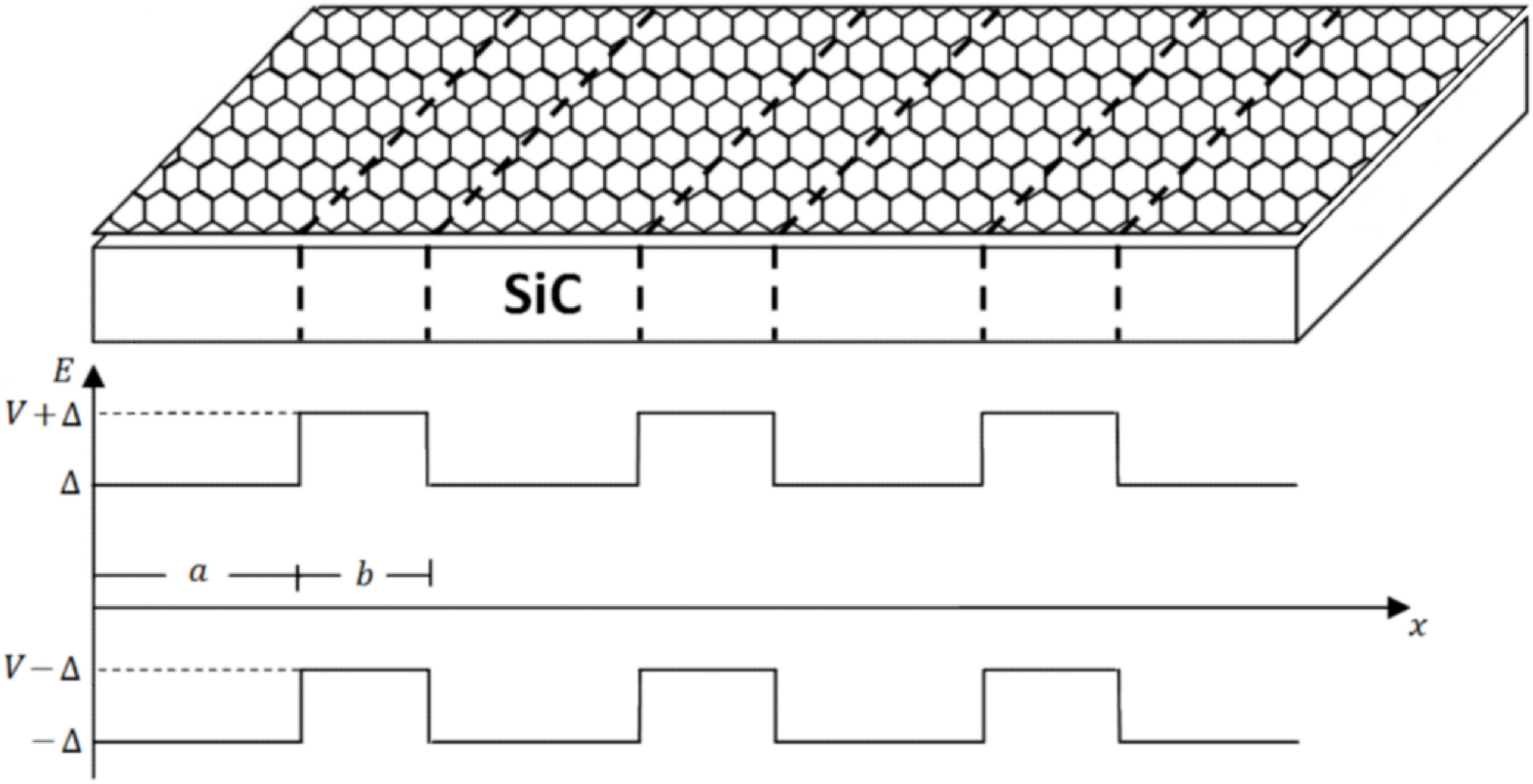}	
\caption{The graphene sheet deposited on a SiC substrate with an external periodic potential $V(x)$ that has a period of $a+b$. The substrate opens a gap of $2\Delta$ in the electronic structure of graphene.}\label{graphene}
\end{figure}

The electronic structure of a graphene sheet in the vicinity of a Dirac point $\textbf{K}$ can be described by an effective Dirac Hamiltonian. Applying an external one-dimensional square-wave potential $V(x)$ and considering an energy gap $2\Delta$ in the electronic structure of graphene, the Dirac-like Hamiltonian reads
\begin{equation}
H=-i\hbar v_F (\sigma_x \partial_x + \sigma_y \partial_y)+V(x)\hat{1}+\Delta\sigma_z \; ,
\end{equation}
where $\sigma_i$ are the Pauli matrices, $\hat{1}$ is the $2\times 2$ unitary matrix and $v_F$ is the Fermi velocity. The energy gap can be realized taking advantage of the influence of the substrate on the electronic properts of graphene. One can, for instance, deposit the graphene sheet in a SiC substrate \cite{SiC}, as shown in Fig. (\ref{graphene}). Other important electronic property of graphene that is affected by the substrate is the Fermi velocity $v_F$. For graphene in different substrates the Fermi velocity has been measured by different authors and their results summarized in \cite{Hwang}. We are considering a periodic potential $V(x)$ with period $a+b$ that is equal to $V$ at $0\leq x< a$ and zero at $a\leq x \leq a+b$.

The Dirac equation is given by
\begin{equation}
H\psi(x,y)=E\psi(x,y) \; ,
\label{dirac}
\end{equation}
where $\psi(x,y)$ is a two-component spinor that represents the two graphene sublattices. Writing
\begin{equation}
\psi(x,y) = e^{ik_y y}\psi(x)
\end{equation}
and replacing in (\ref{dirac}), one will have
\begin{equation}
H=-i\hbar v_F \sigma_x \partial_x + \hbar v_F k_y \sigma_y +V(x)\hat{1}+\Delta\sigma_z \; .
\label{dirac1d}
\end{equation}
Applying the unitary transformation $U_x(\phi/2)=e^{i\frac{\phi}{2}\sigma_x}$, which commutes with $\sigma_x$ and $\hat{1}$ but not with $\sigma_{y,z}$ on can write
\begin{eqnarray}
H^{\prime}&=&U_x H U_x^{\dagger} = -i\hbar v_F \sigma_x \partial_x \nonumber \\
&&+e^{i\frac{\phi}{2}\sigma_x}[\hbar v_F k_y \sigma_y +\Delta \sigma_z]e^{-i\frac{\phi}{2}\sigma_x}+ V(x)\hat{1}
\end{eqnarray}
Using the property $U_j \sigma_i= \sigma_i U_j^{-1}$ if $i \neq j$ we obtain that 
\begin{eqnarray}
e^{i\frac{\phi}{2}\sigma_x}[\hbar v_F k_y \sigma_y &+& \Delta \sigma_z]e^{-i\frac{\phi}{2}\sigma_x}= \nonumber \\
\left[\Delta \cos \phi - \hbar v_F k_y \sin \phi \right] \sigma_z &+& [	\hbar v_F k_y \cos \phi + \Delta \sin \phi] \sigma_y 
\end{eqnarray}
Thus, we can define new effective mass and effective $k_y$ terms
\begin{equation}
\left\{ 
\begin{array}{c c c}
\Delta^*&=&\Delta \cos \phi - \hbar v_F k_y \sin \phi \\
k_y^* &= &\Delta \sin \phi + \hbar v_F k_y \cos \phi
\end{array}
\right.
\end{equation}
Now, defining $\Delta /(\hbar v_F k_y) = \tan \phi$ we make $\Delta^* =0$. So, the Hamiltonian (\ref{dirac1d}) is reduced to
\begin{equation}
H^{\prime}=-i\hbar v_F \sigma_x \partial_x +  k_y^* \sigma_y +V(x)\hat{1} \; ,
\end{equation}
which is the two-dimensional massless Dirac equation for graphene with a periodic potential. The equation
\begin{equation}
H^{\prime}\psi^{\prime}(x,y)=E\psi^{\prime}(x,y) 
\end{equation}
was already solved by different methods \cite{Barbier,Arovas} and the dispersion relation is given by
\begin{eqnarray}
\cos &&(k_x l)=\cos(k_1^*a)\cos(k_2^*b) \nonumber \\
&&+\frac{(k_y^*)^2+E(V-E)}{\hbar^2 v_F^2k_1^*k_2^*}\sin(k_1^*a)\sin(k_2^*b) \; ,
\label{dispersion1}
\end{eqnarray}
where $k_1^*=([E^2- (k_y^*)^2]/\hbar^2 v_F^2 )^{1/2}$, $k_2^*=([(V-E)^2- (k_y^*)^2]/\hbar^2 v_F^2 )^{1/2}$, $k_x$ is the Bloch wave number and we have defined $l=a+b$. In order to  transform back to the original $k_y$ and $\Delta$ terms one can use the inverse transformation
\begin{equation}
\left\{ 
\begin{array}{c c c}
\Delta&=&\Delta^* \cos \phi +  k_y^* \sin \phi \\
\hbar v_F k_y &= &-\Delta^* \sin \phi +  k_y^* \cos \phi
\end{array}
\right.
\end{equation}
As we define $\Delta /(\hbar v_F k_y) = \tan \phi$, we have that $(k_y^*)^2= \Delta^2 + \hbar^2 v_F^2 k_y^2$. Replacing this in Eq. (\ref{dispersion1}) we obtain that the dispersion relation for a 2D massive Dirac equation with a periodic potential is given by
\begin{eqnarray}
\cos &&(k_x l)=\cos(k_1a)\cos(k_2b) \nonumber \\
&&+\frac{k_y^2\hbar^2 v_F^2+E(V-E)+\Delta^2}{\hbar^2 v_F^2k_1k_2}\sin(k_1a)\sin(k_2b) \; ,
\label{dispersion}
\end{eqnarray}
where $k_1=([E^2-\Delta^2]/\hbar^2 v_F^2 - k_y^2)^{1/2}$ and $k_2=([(V-E)^2-\Delta^2]/\hbar^2 v_F^2 - k_y^2)^{1/2}$. Note that at $V=\Delta=0$ we recover the linear dispersion relation of a graphene sheet.

The left hand side of Eq. (\ref{dispersion}) is limited to the interval (-1,1). Therefore, in the right hand side one has allowed and forbidden values for the energy, which implies in the appearance of energy bands with gaps. 

\section{The Electronic Structure}

Having obtained the dispersion relation, in this section we will analyze the electronic structure. In what follows, we shall consider a constant period of the superlattice equal to $60$ nm, i.e., $a+b=60$ nm. As Eq. (\ref{dispersion}) is invariant under simultaneous replacements $E\rightarrow -E$ and $V\rightarrow -V$, only non-negative values of $V$ will be considered. We shall concentrate our discussion on the valence and conductance minibands only, assuming the Fermi level to be in between at any value of V.

\subsection{Band structure with equal well and barrier widths}

\begin{figure}[hpt]
\centering
\includegraphics[width=5cm,height=10cm]{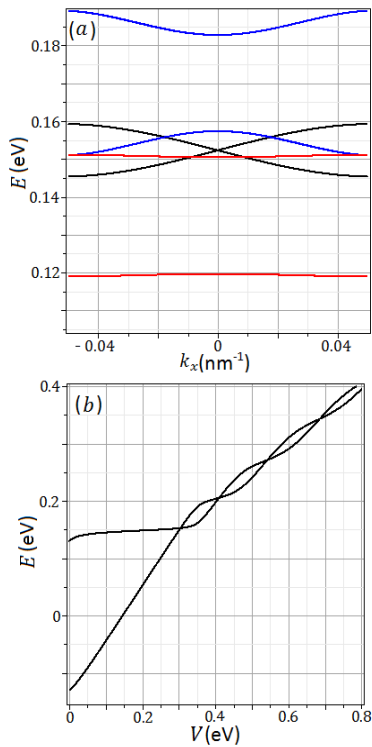}	
\caption{The dispersion relation (\ref{dispersion}) with $a=b=30$ nm. $(a)$ The electron and hole energies as a function of $k_x$, with $k_y = 0$ and $\Delta = 0.13$ eV at $V=270$ meV (red), $V=304.52$ meV (black) and $V=340$ meV (blue), revealing the possibility of closing the energy gap with the potential. $(b)$ The electron and hole minibands as a function of $V$ with $k_x=k_y=0$, which shows the oscillation of the energy gap.}
\label{Ka=b}
\end{figure}

Here we will study the electronic band structure at $a=b=30$ nm in the dispersion relation (\ref{dispersion}), which means that the well and barrier have the same width. In Fig. \ref{Ka=b} $(a)$ are plotted the electron and hole energies as a function of $k_x$ with $k_y=0$ and $\Delta=0.13$ eV for $V=270$ meV (red), $V=304.52$ meV (black) and $V=340$ meV (blue). It can be seen that when the potential increases, the electron and hole minibands shift up. However, the shift of the electron miniband is not equal to the shift of the hole miniband, which implies different electron-hole minigaps for different values of $V$, as shown in Fig. \ref{Ka=b} $(a)$. One can see that it is possible to close the minigap, as happens when $V=304.52$ meV (black), showing that is possible to have a gapped or gapless graphene only changing $V$. It is a consequence of having a position dependent potential. If the potential is constant, the electron and hole minibands are shifted equally and the minigap remain the same, regardless of the value of the potential. 

This is more clear when we look to Fig. \ref{Ka=b} $(b)$, where the electron and hole energies are plotted as a function of $V$ with $k_x=k_y=0$. It should be noted that the contact point is obtained in $k_x=0$, as can be seen in Fig. \ref{Ka=b} $(a)$. Therefore, Fig. \ref{Ka=b} $(b)$ is showing how the electron-hole minigap changes with the potential for $k_y=0$. It can be seen that the minigap oscillates when $V$ changes, and may be zero. For the values of the parameters chosen here, the first value of $V$ that closes the gap is $V=304.52$ meV. The highest value for the energy gap is obtained at $V=0$, which is equal to $2\Delta$. Thus, it is not possible to increase the initial gap in graphene with a periodic potential. It means that a periodic potential can tune the Dirac gap $E_g$ only in the range $0\leq E_g \leq 2\Delta$. So, if $\Delta = 0$, the potential is not able to open a gap.

\begin{figure}[hpt]
\centering
\includegraphics[width=4.5cm,height=4.05cm]{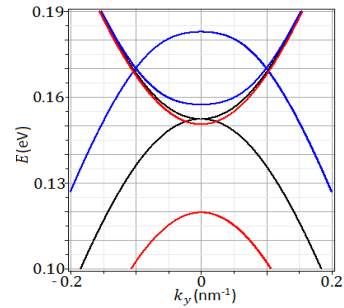}	
\caption{The electron and hole energies as a function of $k_y$ with $a=b=30$ nm, $k_x=0$ and $\Delta=0.13$ eV at $V=270$ meV (red), $V=304.52$ meV (black) and $V=340$ meV (blue). Extra Dirac points appear at $k_y \neq 0$ when $V$ exceeds a critical value $V_c$. In this case $V_c = 304.52$ meV, and the energy gap remains closed.}
\label{kya=b}
\end{figure} 

The electron and hole energies as a function of $k_y$ with $k_x=0$ are plotted in Fig. \ref{kya=b} for the same values of $V$ as in Fig. \ref{Ka=b} $(a)$. When $V=270$ meV (red) and $V=304.52$ meV (black) the minibands have the same behavior that in Fig. \ref{Ka=b} $(a)$, but are narrower. For $V=340$ meV (blue) the minigap at $k_y=0$ opens, however there are extra Dirac points appearing at different values of $k_y$. These extra Dirac points appear when $V$ exceeds a critical value, that is $V_c=304.52$ for the values of the parameters chosen here, and do not disappear. So, from $V=V_c$, the gapped graphene becomes gapless.

In order to find an expression for $V_c$ in terms of the system parameters, let us first localize the contact points in \textbf{k} space. Taking into account the implicit function theorem, one can conclude that at the contact points, where there is an intersection of the bands, the gradient (Jacobian) of the dispersion relation should be zero. Note that $k_1 = k_2$ when $E=E_0=V/2$ and that the contact points are all at $k_x=0$. So, the Eq. (\ref{dispersion}) with $a=b$, $k_x=0$ and $E=E_0$ is given by
\begin{equation}
1=\cos^2(k_1a)+\frac{k_y^2\hbar^2 v_F^2+E_0(V-E_0)+\Delta^2}{\hbar^2 v_F^2k_1^2}\sin^2(k_1a) \; ,
\label{e0}
\end{equation}
which is satisfied when $k_1=n\pi/a$, where $n$ is an integer different of zero, because $n=0$ implies $k_1=0$, which makes the denominator $\hbar^2 v_F^2k_1^2$ in Eq. (\ref{e0}) vanishes. This condition leads to
\begin{equation}
k_y =k_{y_n}=\sqrt{\frac{E_0^2-\Delta^2}{\hbar^2v_F^2}-\left(\frac{n\pi}{a}\right)^2} \; ,
\label{kyn}
\end{equation}
which gives the values of $k_y$ where the contact points are located. The exact location of the contact points are $(E,k_x,k_y)=(E_0,0,k_{y_n})$. Should be remembered that the Dirac points appear only after a critical value of $V=V_c$. The zeros of the equation above give the contact points at $k_y=0$. So, one can write
\begin{equation}
V_n=2\sqrt{\left(\frac{n\pi \hbar v}{a}\right)^2+\Delta^2} \; ,
\end{equation}
which gives the values of $V$ where there is a contact point in Fig. \ref{Ka=b} $(b)$. The critical potential is given by $V_c=V_1$.

The number of contact points can be found from Eq. (\ref{kyn}). When $(E_0^2-\Delta^2)^{1/2}a/\hbar v_F \pi$ is not an integer, the number of contact points is given by
\begin{equation}
N_D=2\left[\frac{a\sqrt{E_0^2-\Delta^2}}{\pi \hbar v_f}\right] \;,
\end{equation}
where $[\cdot \cdot \cdot]$ denotes an integer part. When $(E_0^2-\Delta^2)/\hbar^2 v_F^2 = (n\pi/a)^2$, the number of Dirac points is $N_D=2n-1$. A different way to obtain the number of Dirac points is: when $V_n<V<V_{n+1}$, $N_D=2n$, whereas when $V=V_n$, $N_D=2n-1$. 

\subsection{Band structure with unequal well and barrier widths}

\begin{figure}[hpt]
\centering
\includegraphics[width=5cm,height=10cm]{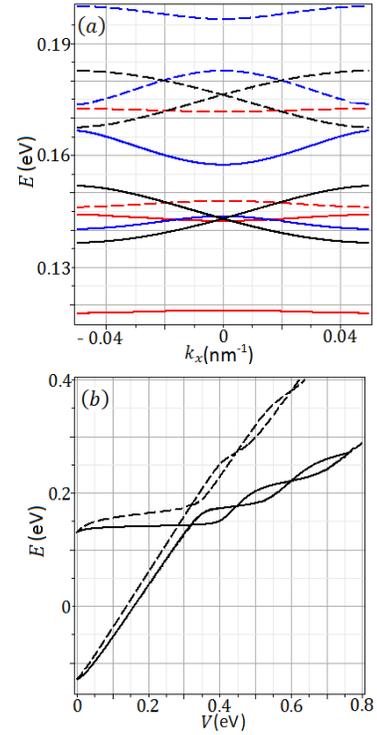}	
\caption{The dispersion relation (\ref{dispersion}) with $a\neq b$. The dashed lines are the minibands with $a=20$ nm and $b=40$ nm, whereas the continuum lines are the minibands with $a=40$ nm and $b=20$ nm. $(a)$ The energy in terms of $k_x$ with $k_y=0$ and $\Delta = 0.13$ eV at $V=290$ meV (red), $V=319.07$ meV (black) and $V=340$ meV (blue). $(b)$ The electron and hole minibands as a function of $V$ with $k_x=k_y=0$.}
\label{Kadiffb}
\end{figure}

Now let us consider the case with $a\neq b$. The dashed lines in Fig. \ref{Kadiffb} are the minibands with $a=20$ nm and $b=40$ nm, whereas the continuum lines represent the minibands with $a=40$ nm and $b=20$ nm. In Fig. \ref{Kadiffb}$(a)$ are plotted the electron and hole minibands as a function of $k_x$ with $k_y=0$ at $V=290$meV (red), $V=319.07$meV (black) and $V=340$meV (blue). As in the case with $a=b$, for different values of $V$ there are different electron-hole minigaps, which may be zero. The oscillation of the minigap is shown in Fig. \ref{Kadiffb} $(b)$, where the minibands as a function of $V$ with $k_x=k_y=0$ are plotted.

It can be seen that the values of $V$ that close the gap when $a=c_1$ and $b=c_2$ are the same for $a=c_2$ and $b=c_1$. However, due to the fact that the potential shifts the minibands, when the graphene region with $V(x)=V$ is wider than the region with $V(x)=0$, there is a larger shift of the minibands. It explains the difference in energy between the dashed and continuum lines in Fig. \ref{Kadiffb}.

\begin{figure}[hpt]
\centering
\includegraphics[width=5cm,height=10cm]{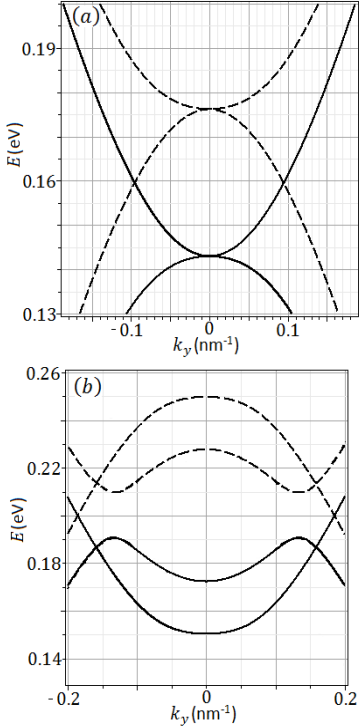}	
\caption{The electron and hole minibands as a function of $k_y$ with $\Delta =0.13$ eV and $k_x=0$ at $(a)$ $V=319.07$ meV and $(b)$ $V=400$ meV. The dashed lines are the minibands with $a=20$ nm and $b=40$ nm, whereas the continuum lines are the minibands with $a=40$ nm and $b=20$ nm. The extra Dirac points at $k_y\neq 0$ are shifted up (down) the Fermi level when $a>b$ ($a<b$).}
\label{kyadifb}
\end{figure} 

In Fig. \ref{kyadifb} we plotted the minibands as a function of $k_y$ with $k_x=0$. Once more, the dashed lines are the minibands with $a=20$ nm and $b=40$ nm, whereas the continuum lines are the minibands with $a=40$ nm and $b=20$ nm. In Fig. \ref{kyadifb} $(a)$ we recovered the first time that the minigap closes at $V=319.07$ meV. In Fig. \ref{kyadifb} $(b)$ we have $V=400$ meV. One can see that, when $a\neq b$, the extra contact points that appears at $k_y\neq 0$ are not in the Fermi level. When $a>b$ $(a<b)$ the contact points are shifted up (down) the Fermi level.

In order to localize the contact points, again, we take advantage of the implicit function theorem. The gradient of the dispersion relation will be zero only if $\sin k_1a = \sin k_2b=0$ and $\cos k_1a=\cos k_2b = \pm 1$. So, one can write
\begin{equation}
k_1 a = \left(\frac{E^2-\Delta^2}{\hbar^2 v_F^2} - k_y^2\right)^{\frac{1}{2}}a = m\pi 
\label{ka}
\end{equation}
and
\begin{equation}
k_2 b =\left(\frac{(V-E)^2-\Delta^2}{\hbar^2 v_F^2} - k_y^2\right)^{\frac{1}{2}}b = m\pi \; ,
\label{kb}
\end{equation}
where $m$ is an integer. Subtracting (\ref{kb}) from (\ref{ka}), one gets
\begin{equation}
E=E_{m}=\frac{V}{2}+\frac{\pi^2 \hbar^2 v_F^2}{2V}\left(\frac{m^2}{a^2}-\frac{m^2}{b^2}\right) \; .
\label{En}
\end{equation}
Replacing the equation above in Eq. (\ref{ka}) one obtains
\begin{equation}
k_{y_m}=\pm \sqrt{\frac{E^2_{m}-\Delta^2}{\hbar^2 v_F^2}-\frac{m^2\pi^2}{a^2}} \; .
\end{equation}
From the zeros of equation above one obtains,
\begin{equation}
V_m=\sqrt{\left(\frac{m\pi \hbar v}{a}\right)^2+\Delta^2}+\sqrt{\left(\frac{m\pi \hbar v}{b}\right)^2+\Delta^2} \; ,
\label{Vn}
\end{equation}
which is the values of $V$ where there is a contact point in Fig. \ref{Kadiffb} $b$. Again, the critical potential, where the graphene superlattice becomes gapless, is $V_c=V_1$. One can see that increasing the difference between $a$ and $b$ the value of $V_c$ increases, as well. Therefore, for a particular $V$, there is always a value of $a$ and $b$ which the graphene is gapped. When $a=b$ in Eqs. (\ref{En})-(\ref{Vn}), the results obtained in the last section are recovered.

\begin{figure}[hpt]
\centering
\includegraphics[width=5cm,height=5cm]{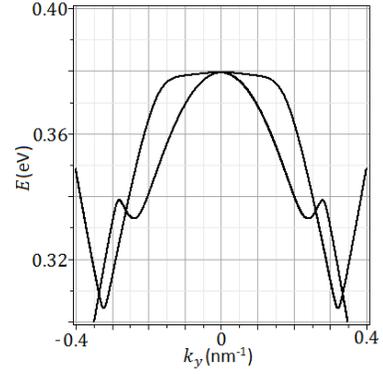}	
\caption{The five Dirac points in the $\textbf{k}$ space with $a=20$ nm, $b=40$ nm and $V=V_3=600.1$ meV. The Dirac points at $k_y \neq 0$ are located below the Fermi energy.}
\label{dp}
\end{figure}

The exact location of the contact points when $V=V_m$ or $V_m<V<V_{m+1}$ is given by
\begin{equation}
(E,k_y)=(E_m,k_{y_m}),(E_{m-1},k_{y_{m-1}}),...,(E_1,k_{y_1}) \; ,
\end{equation}
where $(E_m,k_{y_m})$ is the contact point nearest to $k_y=0$, whereas $(E_1,k_{y_1})$ is the contact point farthest to $k_y=0$. Remember that in our system all contact points are at $k_x=0$. An example is given in Fig. \ref{dp}, where the electron and hole energies are plotted as a function of $k_y$ with $a=20$ nm, $b=40$ nm and $V=V_3=600.1$ meV. The location of the contact point nearest to $k_y=0$ is $(E_3,0)$. The next nearest contact point is $(E_2,k_{y_2})$ and the farthest contact point is located at $(E_1,k_{y_1})$.

The number of contact points is the same obtained in the last section with $a=b$.

\subsection{Dispersion relation near the contact points}

Now, let us analyze the electronic structure near the contact points. The electronic structure in the vicinity of the contact points has been studied in the case of a gapless graphene superlattice \cite{Barbier} and for a graphene superlattice with spatially modulated gap \cite{PhysRevB.86.205422}. In both cases, the discussion was restricted to the particular case $a=b$. For the sake of comparison, we will consider the special case with $a=b$ and then consider the general case with $a\neq b$. For this purpose, one has to expand Eq. (\ref{dispersion}) in the vicinity of the contact points obtained in the last sections. 

Let us first consider the special case $a=b$. In order to obtain the behavior of all contact points, one has to consider the contact points at $E=E_0(V_n)$, $k_x=0$ and $k_y=0$ and the contact points at $E=E_0(V_n)$, $k_x=0$ and $k_y=k_{y_n}$. Thus, expanding Eq. (\ref{dispersion}) into the Taylor series up to second order of $E-E_0(V_n)$, $k_x$ and $k_y^2$, one gets
\begin{eqnarray}
\varepsilon_1 =&\pm &\left[\frac{k_y^4l^4(d^2 u^2 - 4d^4)}{k_y^2l^2(2u^4-8d^4)+4\pi^2u^6-16\pi^2d^2u^4}\right. \nonumber \\
 &+&\left. \frac{k_x^2\pi^2 l^2(192d^4u^2-256d^6+4u^6-48u^4d^2)}{k_y^2l^2(2u^4-8d^4)+4\pi^2u^6-16\pi^2d^2u^4}\right]^{1/2} \; ,
\label{E1}
\end{eqnarray}
where we have defined $\varepsilon_1 = (E-E_0)l/\hbar v_F$, $u=V_nl/4\pi\hbar v_F$ and $d=\Delta l/4\pi\hbar v_F$. The positive and negative signs represent the electron and hole minibands, respectively. Writing $\overline{k}_y=k_y - k_{y_n}$ and expanding Eq. (\ref{dispersion}) up to the lowest order of $E-E_0$, $k_x$ and $\overline{k}_y$, one obtains the dispersion law in the vicinity of the contact points located at $k_y\neq 0$, which is given by
\begin{equation}
\varepsilon_2 = \pm \sqrt{k_x^2l^2\frac{n^4}{u^4}+\overline{k}_y^2l^2\left(1-2\frac{n^2}{u^2}+4\frac{d^2n^2}{u^4}-4\frac{d^2}{u^2}+\frac{n^4}{u^4}\right)}\;.
\label{E2}
\end{equation}
It should be noted that, when $a=b$, the electron and hole minibands are symmetric related to $\varepsilon_i=0$, which does not happen when there is a periodic modulation of the energy gap in graphene \cite{PhysRevB.86.205422}.

\begin{figure}[hpt]
\centering
\includegraphics[width=8cm,height=8cm]{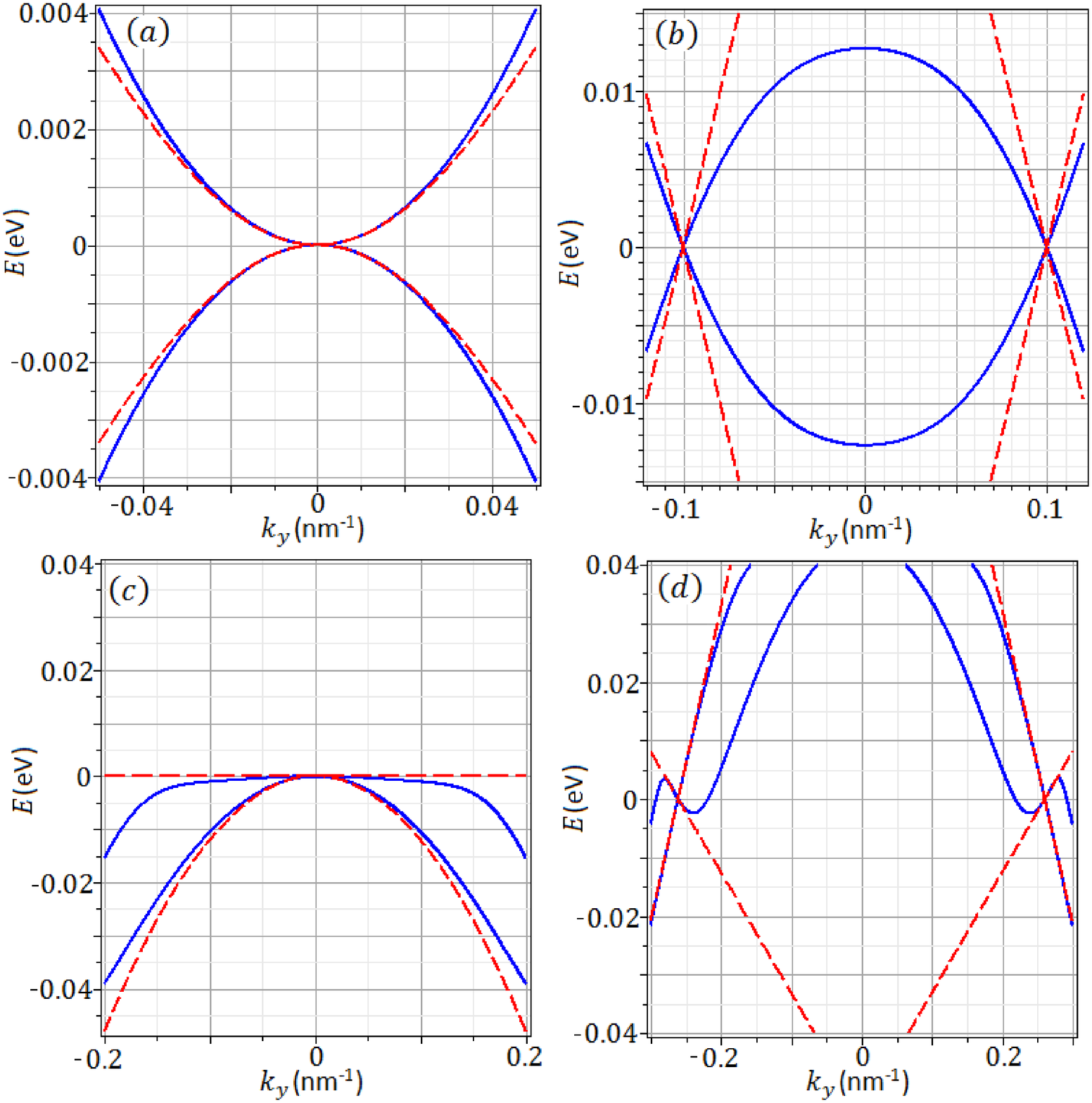}	
\caption{The electron and hole energies in terms of $k_y$ with $k_x=0$ in order to compare the exact dispersion relation (blue line) with the spectrum obtained around the contact points (red dashed line). $(a)$ $\varepsilon_1$ at $V=V_1$. $(b)$ $\varepsilon_2$ at $V=340$ meV and $n=1$. $(c)$ $\varepsilon_3$ at $V=600.1$ meV and $m=3$. $(d)$ $\varepsilon_4$ at $V=600.1$ meV and $m=2$.}
\label{drsmalle}
\end{figure}

In Fig. \ref{drsmalle} $(a)$ we plotted $\varepsilon_1$ (dashed red line) with $V=V_1$ and $k_x=0$ as a function of $k_y$ and compare with the exact dispersion relation (\ref{dispersion}) (blue line). One can see that the expansion (\ref{E1}) is good in the vicinity of the contact point. Expanding $\varepsilon_1$ in powers of $k_y$ with $k_x=0$, one gets
\begin{equation}
\varepsilon_1 = \pm \frac{d}{2\pi u^2}k_y^2 + O(k_y^4) \; ,
\end{equation}
which gives a parabolic electron and hole minibands, as can be seen in Fig. \ref{drsmalle} $(a)$, in contrast to the conical dispersion around the original Dirac point in a gapless graphene. In the limit when $u^2\gg d$, the dispersion along $k_y$ becomes flat. Note that in the case of a gapless graphene superlattice \cite{Barbier}, the dispersion along $k_y$ is given by $\varepsilon_1 \sim \pm k_y^3$. In Fig. \ref{drsmalle} $(b)$ we compare $\varepsilon_2$ (dashed red line) with Eq. (\ref{dispersion}) (blue line) at $V=340$ meV. Again, there is an agreement between the exact dispersion relation and the expansion (\ref{E2}) near the contact point. However, the dispersion is linear along $k_y$, which does not happen in the contact points located at $k_y=0$. 

Considering now the most general case with $a\neq b$, one can expand Eq. (\ref{dispersion}) with $V=V_m$ up to the lowest order of $E-E_m$, $k_x$ and $k_y^2$, and obtain the dispersion relation in the vicinity of the contact points at $k_x=k_y=0$ and $E=E_m$, that is given by
\begin{equation}
\varepsilon_3 = \alpha_1 k_y^2 \pm \sqrt{\beta_1 k_y^4 + \gamma_1 k_x^2}\; .
\label{E3}
\end{equation}
The dispersion near the contact points at $k_y\neq 0$ can be written as
\begin{equation}
\varepsilon_4 = \alpha_2 \overline{k}_y \pm \sqrt{\beta_2 \overline{k}_y^2 + \gamma_2 k_x^2} \; ,
\label{E4}
\end{equation}
where $\overline{k}_y = k_y - k_{y_m} $. The coefficients $\alpha_1$, $\beta_i$ and $\gamma_i$, with $i=1,2$, depend on $a$, $b$, $V$ and $\Delta$. They are too large to be write down here. When $a=b$, the coefficients $\alpha_1$ and $\beta_1$ vanish and $\varepsilon_3$ becomes $\pm k_x(V_n^2 - 4\Delta^4)/V_n^2$. For this reason, it was necessary to expand Eq. (\ref{dispersion}) up to second order of $E-E_0(V_n)$, $k_x$ and $k_y^2$ to get $\varepsilon_1$.

One can note that, in contrast to the case with equal well and barrier widths, when $a\neq b$ the electron and hole minibands are not symmetric related to $\varepsilon_i =0$ due to the coefficient $\alpha_i$. However, as in the case with $a=b$, the minibands along the $k_y$ direction are parabolic in the contact points at $k_y =0$ and conical at $k_y \neq 0$. In Fig. \ref{drsmalle} $(a)$ and $(b)$ we compare $\varepsilon_3$ and $\varepsilon_4$, respectively, with the exact dispersion relation (\ref{dispersion}). We consider $V=V_3=600.1$ meV and plotted $\varepsilon_4$ in the vicinity of the contact points located at $(E_2,k_{y_2})$. 

Should be mentioned that the dispersion relation is linear along $k_x$ around all contact points. So, the energy surface $\varepsilon(k_x,k_y)$ is conical in the vicinity of the contact points at $k_y\neq 0$ and has a lenslike shape around the contact points at $k_y=0$.

\subsection{Group velocity around the contact points}

Let us now use the spectrum for small energies obtained above to find the effective Fermi velocity around the contact points. The components of the velocity in the vicinity of the contact points are given by $v_{x_i}/v_F=\partial \varepsilon_i/\partial k_x$ and $v_{y_i}/v_F=\partial \varepsilon_i/\partial k_y$, where $i=1,2,3,4$ denote the four kinds of contact points. The expressions for the components of the velocity can be seen in the Appendix.

\begin{figure}[hpt]
\centering
\includegraphics[width=8cm,height=12cm]{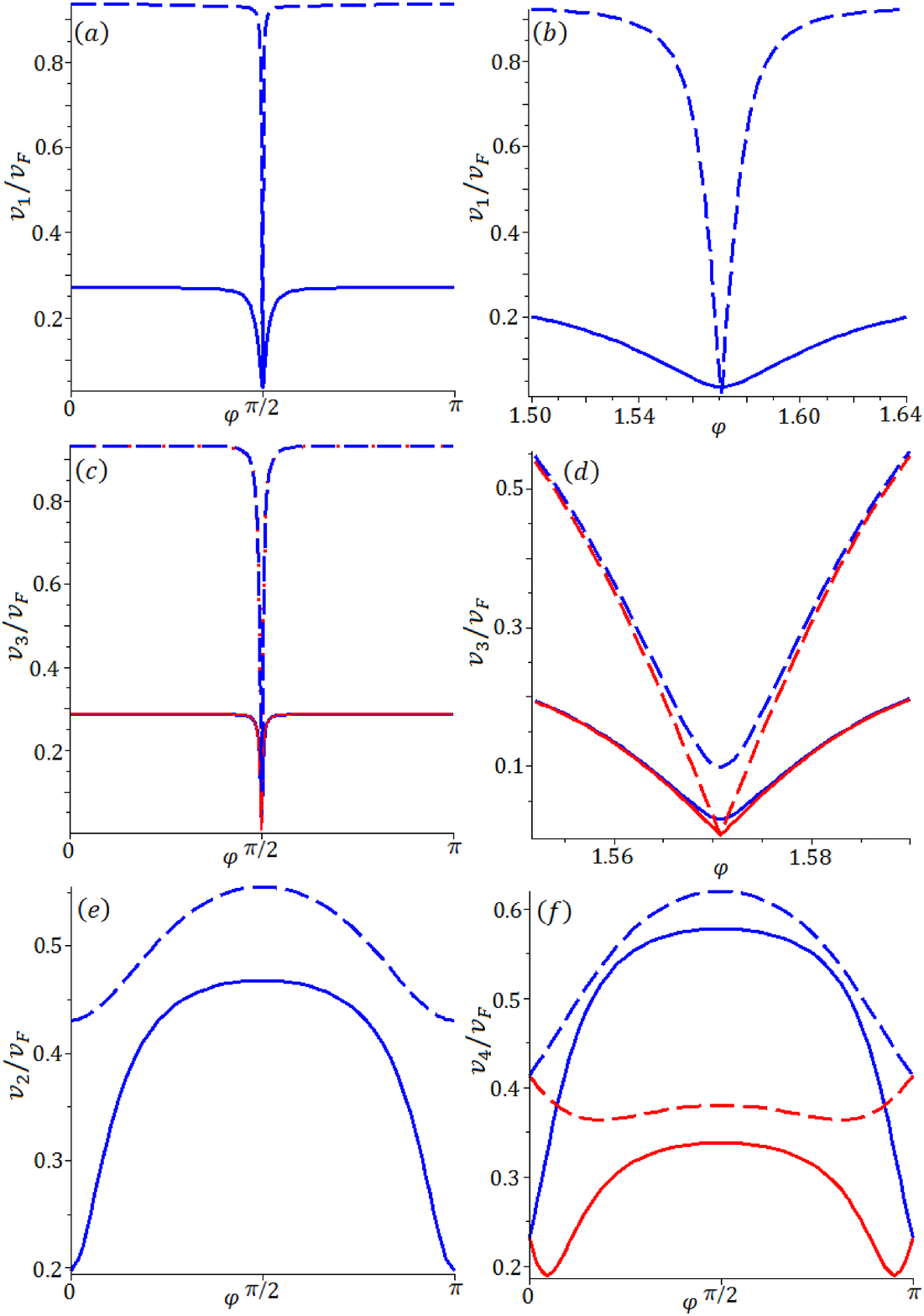}	
\caption{The electron (blue) and hole (red) group velocities in the vicinity of the four kinds of contact points as a function of $\varphi$ at $\Delta=0.13$ eV (continuum line) and $\Delta=26.5$ meV (dashed line). $(a)$ and $(b)$ $v_1/v_F$ at $V_{cont}=304.52$ meV and $V_{dash}=212.03$ meV. $(e)$ $v_2/v_F$ at $V_{cont}=357$ meV and $V_{dash}=313$ meV. In these cases, the electron and hole velocities are equal. $(c)$ and $(d)$ $v_3/v_F$ with $a=20$ and $b=40$ nm at $V_{cont}=319.07$ meV and $V_{dash}=237.65$ meV. $(f)$ $v_4/v_F$ with $a=20$ nm and $b=40$ nm at $V_{cont}=383.5$ meV and $V_{dash}=351.48$ meV. $V_{cont}$ and $V_{dash}$ are the value of $V$ for the continuum and dashed lines, respectively.}  
\label{velo}
\end{figure}

The anisotropy of the electron and hole velocities in the $(k_x,k_y)$ plane can be seen clearly if one introduces a polar angle $\varphi$ with the relations $k_x=Q \cos \varphi$ and $k_y=Q \sin \varphi$, where $Q=\sqrt{k_x^2+k_y^2}$. In Fig. \ref{velo} the absolute value of the velocity $v_i=\sqrt{v_{x_i}^2+v_{y_i}^2}$ as a function of $\varphi$ was plotted for two different values of the energy gap: $\Delta=0.13$ eV (continuum line) and $\Delta=26.5$ meV (dashed line), which correspond to the graphene on a SiC and \textit{h}-BN substrate, respectively. It should be mentioned that the Fermi velocity $v_F$ in graphene on these two substrate is $1.15 \cdot 10^6$ m/s and $1.49 \cdot 10^6$ m/s, respectively \cite{Hwang}. One can see that the velocity is sensitive to the energy gap $\Delta$. The velocity has smaller values for larger values of $\Delta$.

In Fig. \ref{velo} $(a)$ and $(e)$ we plotted the velocity for $a=b$ at $V=V_1$ ($V_1=304.52$ meV when $\Delta=0.13$ eV and $V_1=212.03$ meV when $\Delta=26.5$ meV) and at the intermediate value between $V_1$ and $V_2$ ($V=357$ meV when $\Delta=0.13$ eV and $V=313$ meV when $\Delta=26.5$ meV), respectively. When $a=b$ the electron and hole velocities are equal, in consequence of the symmetry between the electron and hole minibands. This does not happen in the case with a modulated energy gap \cite{PhysRevB.86.205422}, where the electron and hole velocities are not the same. For small values of $\Delta$, $v_1$ is close to $v_F$ for almost all values of the angle $\varphi$, having a narrow dip in the vicinity of $\varphi=\pi/2$, as can be seen in Fig. \ref{velo} $(a)$. This is due to the fact that $v_{x_1}$ is much greater than $v_{y_1}$ for all angle $\varphi$ except in the vicinity of $\pi/2$, where both $v_{x_1}$ and $v_{y_1}$ are small, in consequence of the lenslike shape of the energy surface. When $\Delta$ increase the velocity decrease and the value of $v_1$ become much smaller than $v_F$. The dip remains at $\varphi=\pi/2$, but it become a little wider. A similar behavior was obtained in \cite{PhysRevB.86.205422}. In Fig. \ref{velo} $(b)$ there is a zoom of the dip region. Different of $\varepsilon_1$, the energy surface generated by $\varepsilon_2$ is conical, but it is not an isotropic cone, generating an anisotropy in the velocity. In this case, with a small energy gap, the electron and hole velocities have only a little variation around $v_F/2$, as can be seen in Fig. \ref{velo} $(e)$. Increasing $\Delta$, there is a stronger anisotropy, in contrast with \cite{PhysRevB.86.205422}.

The electron and hole velocities for $a=20$ nm and $b=40$ nm at $V=V_1$ ($V_1=319.07$ meV when $\Delta=0.13$ eV and $V_1=237.65$ meV when $\Delta=26.5$ meV) and at the intermediate value between $V_1$ and $V_2$ ($V=383.5$ meV when $\Delta=0.13$ eV and $V=351.48$ meV when $\Delta=26.5$ meV) are plotted respectively in Fig. \ref{velo} $(c)$ and $(f)$. When $a\neq b$ the electrons and hole minibands are asymmetric, so the electron and hole velocities are not equal. In Fig. \ref{velo} $(c)$ we consider the contact point at $k_y=0$. The behavior of the velocity in this case is very similar with the case with $a=b$. The main difference is that the dip has a width slightly different. One can note that the electron and hole velocities are almost the same, differing slightly in the vicinity of $\varphi=\pi/2$. The Fig. \ref{velo} $(d)$ is an extension of the dip region. The energy surface generated by $\varepsilon_4$ is conical, but is a tilted and not isotropic cone. In Fig. \ref{velo} $(f)$ we plotted $v_4$. It can be seen that the electron and hole velocities are equal at $\varphi=0,\pi$ and differ widely for other values of $\varphi$. As in the case with $a=b$, when $\Delta$ increases the anisotropy of the velocity becomes greater.

\section{Conclusions}

We have analyzed the electronic structure of a gapped graphene superlattice with a piecewise constant periodic potential using the continuum model based on an effective Dirac equation. We consider that the energy gap is generated by an appropriate substrate, which changes the Fermi velocity, as well.

It was shown that the energy gap oscillates when the potential $V$ changes continuously at $k_x=k_y=0$ and is zero at discrete values $V_n$. When $V>V_1$, extra Dirac points appear at $k_y\neq 0$ and never disappear. Thus, beginning with a critical potential $V_c=V_1$, the graphene system becomes gapless. In the special case of equal well and barrier widths, these extra Dirac points are located in the Fermi level and the electron and hole minibands are symmetric, whereas with an unequal well and barrier widths the extra Dirac points are no longer at the Fermi level and the electron and hole minibands are asymmetric. We found that if the initial energy gap $E_g$ in graphene is equal to $2\Delta$, it is possible to tune the energy gap with a periodic potential in the range $0\leq E_g \leq 2\Delta$. We found the locations of all contact points and it was shown that the greater the difference between the well and barrier width, the greater the critical potential $V_c$. Finally, we obtained the dispersion relation in the vicinity of all contact points and used it to find the effective group velocity of the carriers. The velocity has a strong anisotropy around the contact points and is sensitive to the energy gap. Extra Dirac points of different kinds have been already studied in graphene superlattices. Analyzing the electronic structure near the contact points, we showed that the extra Dirac points obtained here have a different behavior compared to previously studied. The results obtained here can be used in the fabrication of graphene--based devices. 

\begin{acknowledgments}
I thank M. A. H. Vozmediano for helping me to revise and correct mistakes in a previous version of the manuscript. This work was partially supported by CNPq and CNPq-MICINN binational. 
\end{acknowledgments}

\appendix*
\section{The components of the group velocity around the contact points}

The components of the velocity in the vicinity of the contact points are given by $v_{x_i}/v_F=\partial \varepsilon_i/\partial k_x$ and $v_{y_i}/v_F=\partial \varepsilon_i/\partial k_y$, with $i=1,2,3,4$. Thus,
\begin{equation}
\frac{v_{x_1}}{v_F}= \pm \frac{k_x B}{\sqrt{(k_y^4 A + k_x^2 B)(k_y^2 C + D)}} \; ,
\label{vx1}
\end{equation}
\begin{equation}
\frac{v_{y_1}}{v_F}=\pm \frac{k_y^5 AC +2k_y^3AD-k_yk_x^2BC}{\sqrt{(k_y^4 A + k_x^2 B)(k_y^2 C + D)^3}} \; ,
\label{vy1}
\end{equation}
\begin{equation}
\frac{v_{x_2}}{v_F}=\pm \frac{k_x n^4/u^4}{\sqrt{k_x^2\frac{n^4}{u^4}+\overline{k}_y^2\left(1-2\frac{n^2}{u^2}+4\frac{d^2n^2}{u^4}-4\frac{d^2}{u^2}+\frac{n^4}{u^4}\right)}} \; ,
\label{vx2}
\end{equation}
\begin{equation}
\frac{v_{y_2}}{v_F}=\pm \frac{\overline{k}_y \left(1-2\frac{n^2}{u^2}+4\frac{d^2n^2}{u^4}-4\frac{d^2}{u^2}+\frac{n^4}{u^4}\right)}{\sqrt{k_x^2\frac{n^4}{u^4}+\overline{k}_y^2\left(1-2\frac{n^2}{u^2}+4\frac{d^2n^2}{u^4}-4\frac{d^2}{u^2}+\frac{n^4}{u^4}\right)}} \; ,
\label{vy2}
\end{equation}
\begin{equation}
\frac{v_{x_3}}{v_F}=\pm \frac{k_x \gamma_1}{\sqrt{\beta_1 k_y^4 + \gamma_1 k_x^2}} \; ,
\label{vx3}
\end{equation}
\begin{equation}
\frac{v_{y_3}}{v_F}=2\alpha_1 k_y \pm \frac{2\beta_1 k_y^3}{\sqrt{\beta_1 k_y^4 + \gamma_1 k_x^2}}\; ,
\label{vy3}
\end{equation}
\begin{equation}
\frac{v_{x_4}}{v_F}=\pm \frac{\gamma_2 k_x}{\sqrt{\beta_2 \overline{k}_y^2 + \gamma_2 k_x^2}}
\label{vx4}
\end{equation}
and
\begin{equation}
\frac{v_{y_4}}{v_F}=\alpha_2 \pm \frac{\beta_2 \overline{k}_y}{\sqrt{\beta_2 \overline{k}_y^2 + \gamma_2 k_x^2}}\; ,
\label{vy4}
\end{equation}
where we define
\begin{equation}
A=l^2(d^2u^2-4d^4)\;,
\end{equation}
\begin{equation}
B=\pi^2(192d^4u^2-256d^6+4u^6-48u^4d^2)\;,
\end{equation}
\begin{equation}
C=l^2(2u^4-8d^4)
\end{equation}
and
\begin{equation}
D=4\pi^2u^4(u^2-4d^2)\;.
\end{equation}

\end{document}